\documentclass[twocolumn,english,aps,pra,eqsecnum]{revtex4-1}
\usepackage[T1]{fontenc}
\usepackage[latin9]{inputenc}
\usepackage{color}
\usepackage{amsmath}
\usepackage{amssymb}
\usepackage{graphicx}
\usepackage{esint}

\makeatletter
\@ifundefined{definecolor}
 {\usepackage{color}}{}
\@ifundefined{definecolor}{\usepackage{color}}{}

\usepackage[english]{babel}

\makeatother

\usepackage{babel}
\begin{document}

\title{Signifying the Schrodinger cat in the context of testing macroscopic
realism }

\author{M. D. Reid$^{2}$}

\affiliation{$^{2}$Centre for Quantum and Optical Science, Swinburne University
of Technology, Melbourne 3122, Australia}
\begin{abstract}
Macroscopic realism (MR) specifies that where a system can be found
in one of two macroscopically distinguishable states (a cat being
dead or alive), the system is always predetermined to be in one or
other of the two states (prior to measurement).   Proposals
to test MR generally introduce a second premise to further qualify
the meaning of MR.  This paper examines two such models, the first
where the second premise is that the macroscopically distinguishable
states are quantum states (MQS) and the second where the macroscopcially
distinguishable states are localised hidden variable states (LMHVS).
We point out that in each case in order to negate the model, it is
necessary to assume that the predetermined states give microscopic
detail for predictions of measurements. Thus, it is argued that
many cat-signatures do not negate MR but could be explained by microscopic
effects such as a photon-pair nonlocality. Finally, we consider
a third model, macroscopic local realism (MLR), where the second premise
is that measurements at one location cannot cause an instantaneous
macroscopic change to the system at another. By considering amplification
of the quantum noise level via a measurement process, we discuss
how negation of MLR may be possible.
\end{abstract}
\maketitle

\section{Introduction}

In his essay of 1935, Schrodinger considered the quantum interaction
of a microscopic system with a macroscopic system \cite{Schrodinger-1}.
After the interaction, the two systems become entangled. If the macroscopic
system were likened to a cat, then according to the standard interpretation
of quantum mechanics, it would seem possible for the cat to be in
a state that is \emph{neither dead nor alive}. The ``Schrodinger
cat-state'' can take many different forms, depending on the particular
realisation employed for the microscopic and macroscopic systems and
their interaction \cite{cats,cats2,catsphil,svetcats,ystbeamsplit-1-1,lgcats,mirrorcats}.

In this paper, I consider how to experimentally test the \emph{interpretation}
of the Schrodinger cat-state. The quantum state describing the microscopic
and macroscopic systems after the interaction can be written as 
\begin{equation}
|\psi\rangle=\frac{1}{\sqrt{2}}\Bigl(|dead\rangle_{C}|\downarrow\rangle_{S}+|alive\rangle_{C}|\uparrow\rangle_{S}\Bigr)\label{eq:cat}
\end{equation}
Here $|\uparrow\rangle$ and $|\downarrow\rangle$ represent two distinct
states for the microscopic sytsem $S$, and the $|dead\rangle$ and
$|alive\rangle$ symbolise two macroscopically distinct states for
the macroscopic system $C$ (that we will call the ``cat'' or the
``cat-system''). The interpretation of the ``cat'' in the superposition
state (\ref{eq:cat}) is that it is ``neither dead nor alive''.
If the cat-system is a pointer of measurement apparatus that has coupled
to the microscopic spin system, then the interpretation is suggestive
that the pointer is in ``two places at once'' \cite{pointer}. While
different signatures have been proposed for Schrodinger cat states
\cite{catmeasures,macro-coh_verdral,bognoon,eric_marg,eprcat,LG,lgpapers,svet},
they are not all equivalent. The words ``neither dead nor alive''
can be interpreted in different ways. 

The issue of testing the interpretation of the cat-state amounts to
testing the classical premise of ``\emph{macroscopic realism}''
(MR). Leggett and Garg gave a proposal for such a test, in their formuation
of the Leggett-Garg inequalities \cite{LG}. They introduced a framework
for the meaning of MR, which was to consider a system that would always
be found in one of two macroscopically distinguishable states (e.g.
``dead'' or ``alive''). They stated as the premise of MR that
the system is\emph{ always in one or other of these states prior to
measurement}. A hidden variable is introduced, to denote which of
these states the system is in, prior to the measurement. We will denote
this hidden variable by $\lambda_{M}$ and refer to it as the ``\emph{macroscopic
hidden variable}''. 

The objective of this paper is to consider ways to test MR and to
link these tests with signatures of the Schrodinger cat-state. To
do this, we are careful at the outset to clarify the definition of
MR. MR asserts that the result of a measurement $\hat{M}$ that is
used to distinguish whether the cat-system is dead or alive is predetermined.
Because the dead and alive states are macroscopically distinguishable,
the measurement $\hat{M}$ can be made with a very large uncertainty
(lack of resolution in the outcomes) and still be 100\% effective.
This means that in assuming MR, we classify the state of the cat by
the single parameter $\lambda_{M}$ and \emph{do not concern ourselves
with microscopic properties or predictions of that state.} 

In order to provide a workable signature for an experiment, previous
tests of macroscopic realism have introduced a second premise. Once
the second premise is introduced, there is no longer a direct test
of MR, because the signature if verified experimentally can be due
to failure of the second premise, rather than MR. It is essential
therefore that the second premise be as powerful as the assumption
of MR itself. Leggett and Garg introduced the second premise of macroscopic
noninvasive measurability \cite{LG}, which can be difficult to
justify in real experiments and which has motivated various forms
of non-invasive measurement \cite{lgpapers}.

In this paper we examine three alternative approaches. First, in Sections
II and III, we analyse the common methodologies for signifying a Schrodinger
cat state, pointing out that there again a second premise apart from
MR is assumed. Depending on which signature is used, the second premise
is that the macroscopically distinguishable states of the system are
\emph{quantum} states, or else \emph{localised hidden variable} states.
These two different sorts of signatures, that we call Type I and II,
are discussed in Sections II and III. In each case, assumptions are
made about the \emph{microscopic} predictions of those states for
measurements other than $\hat{M}$. This means that the signatures
do not imply negation of MR (as defined by the macroscopic hidden
variable $\lambda_{M}$), but could be explained if we allow that
the cat-system be described by hidden variable states, or else if
we allow that there are microscopic nonlocal effects on the cat-system.
Examples of signatures include  violations of Svetlichny-type inequalities
that reveal genuine multipartite Bell nonlocality for Greenberger-Horne-Zeilinger
(GHZ) states \cite{svet}. 

In the third approach, presented in Section V, we introduce as the
second premise the assumption of \emph{macroscopic locality} (ML).
ML asserts that measurements at one location cannot cause an instantaneous
macroscopic change to the system at another. The combined premises
of MR and ML are called macroscopic local realism (MLR) \cite{MLR,mdrmlr,mdrmlr2}.
A test of MLR can be constructed using Bell inequalities predicted
to hold for two spatially separated cat-systems. We point out that
MLR cannot generally be expected to fail, because of bounds placed
on the predictions of quantum mechanics by the uncertainty relation
\cite{MLR-uncert,gisinuncert}. However, we show such tests become
possible if one considers experiments that as part of the measurement
process provide amplification of the quantum noise level \cite{mdrmlr2,MLR}.
In this case, the meaning of ``macroscopically distinguishable''
refers to particle number differences $\delta$ that are large in
an absolute sense, but small compared to the total number of particles
of the system. The second premise is the necessary co-premise of
MR for the experimental scenario where there are two cat-systems.
Proposed experimental arrangements are based on states that predict
a violation of Bell inequalities for continuous variable measurements
\cite{cvbell,noonbellalex,pair-coh}.

In Section IV, it is explained that the signatures considered in Section
II and III do not allow a direct negation of the macroscopic realism
(MR) i. e. they do not directly falsify the macroscopic hidden variable
$\lambda_{M}$. Logically, the signatures can be realised if the second
premise fails with the first one (MR) upheld. This leaves open the
simplest interpretation of the macroscopic pointer (of the cat-state
(\ref{eq:cat})), that the pointer is located at one position or another
but subject to microscopic nonlocal effects due to the entanglement
with the spin system. By contrast, the tests of Section V are predicted
to reveal mesoscopic nonlocal interactions between two pointers. We
give a discussion of the correlation between these pointers and the
possibility of inferring a ``both worlds'' (that the cat is ``dead
and alive'') interpretation.

\section{Type I cat-signatures: negating macroscopic quantum realism}

We consider a macroscopic or mesoscopic system $C$ (called the ``cat'')
and a measurement $\hat{M}$ on the system that yields binary outcomes.
The outcomes are distinct by a quantifiable amount (referred to as
$N$) and correspond to states that we regard as macroscopically distinct
in the limit $N\rightarrow\infty$. The two outcomes are labelled
``dead'' and ``alive'' for simplicity, though for finite $N$
the outcomes are only ``$N$-scopically distinct''. The outcomes
for the measurement $\hat{M}$ may arise from an observable whose
results are binned into two categories, bin $1$ giving the outcome
``dead'' and bin 2 giving the outcome ``alive''. 

The signature for an ``$N$-scopic cat-state'' is a negation that
the system $C$ can be described as a \emph{classical probabilistic
mixture }of states that are either ``dead'' or ``alive''. For
a Type I signature there is the extra assumption that the ``dead''
and ``alive'' states are necessarily given by a quantum density
operator description. Such classical mixtures\emph{ }can be expressed
as \cite{bognoon} 
\begin{equation}
\rho=P_{1}\rho_{1}+P_{2}\rho_{2}\label{eq:mixtureI}
\end{equation}
Here $\rho_{1}$ is a density operator for the system $C$ giving
a result for measurement $\hat{M}$ in bin $1$ (and is thus a ``dead''
state); and $\rho_{2}$ is a density operator giving a result in bin
$2$ (and is thus an ``alive'' state). The $P_{1}$, $P_{2}$ are
probabilities for the system being in state $\rho_{1}$ or $\rho_{2}$
respectively ($P_{1}+P_{2}=1$). We call the negation of the models
(\ref{eq:mixtureI}) the falsification of \emph{macroscopic quantum
realism.}

The model (\ref{eq:mixtureI}) can be negated given the restrictions
imposed because $\rho$ is a \emph{mixture} of quantum states, and
also because the $\rho_{i}$ are \emph{quantum} density operators.
It is straightforward to find criteria to negate (\ref{eq:mixtureI}).
These criteria, that negate \emph{all} relevant classical mixtures
where the regions $1$ and $2$ suitably defined, provide Type I signatures
of a Schrodinger cat-state. 

To illustrate, let us consider the superposition state
\begin{equation}
|\psi_{N}\rangle=\frac{1}{\sqrt{2}}\Bigl(|N\rangle+e^{i\phi}|0\rangle\Bigr)\label{eq:sup1}
\end{equation}
Here $|n\rangle$ is the eigenstate of mode number $\hat{n}$ with
number eigenvalue $n$ and we let $\hat{M}=\hat{n}$. The binned
regions $1$ (``dead'') and $2$ (``alive'') are those that give
outcomes for $\hat{n}$ as less than $N/2$, or greater than or equal
to $N/2$, respectively (Figure 1). To signify that an experimental
system $C$ cannot be described as a mixture (\ref{eq:mixtureI}),
we proceed as follows: For any model (\ref{eq:mixtureI}), we denote
the mean and variance in the predictions for $\hat{n}$ given the
system is in $\rho_{i}$ by $\langle\hat{n}\rangle_{i}$ and $(\Delta\hat{n})_{i}^{2}$
($i=1,2$). For any mixture (\ref{eq:mixtureI}), the inequality
\begin{equation}
\Bigl(\sum_{i}(\Delta\hat{n})_{i}^{2}\Bigr)(\Delta\hat{P}^{N})^{2}\geq\frac{1}{4}|\langle\hat{C}\rangle|^{2}\label{eq:sigcatone-2-2}
\end{equation}
holds. Here $\hat{C}=\left[\hat{n},\hat{P^{N}}\right]$ and $\hat{P}=(\hat{a}-\hat{a}^{\dagger})/i$
is the mode quadrature amplitude, the\textcolor{black}{{} $\hat{a}^{\dagger}$,
$\hat{a}$ being the creation, destruction operators for the single-mode
system}. The proof is given in Ref. \cite{rynoonpaper} and is based
on the fact that for any observable $\hat{B}$, the mixture (\ref{eq:mixtureI})
implies $(\Delta\hat{B})^{2}\geq\sum_{i}P_{i}(\Delta\hat{B})_{i}^{2}$
where $(\Delta\hat{B})_{i}^{2}$ is the variance of $\hat{B}$ for
the state $\rho_{i}$. It is also necessary to use that each $\rho_{i}$
is a \emph{quantum} state and therefore for two conjugate observables
$\hat{A}$ and $\hat{B}$ such that $\hat{C}=[\hat{A},\hat{B}]$,
the quantum uncertainty relation $\Delta\hat{A}\Delta\hat{B}\geq\frac{1}{2}|\langle\hat{C}\rangle|$
must hold. We see then that the violation of the inequality (\ref{eq:sigcatone-2-2})
is a Type I signature for a cat-state. 

The superposition $|\psi_{N}\rangle$ violates the inequality (\ref{eq:sigcatone-2-2}).
The predicted experimental outputs for $\hat{n}$ are given in Figure
1 which implies $(\Delta\hat{n})_{i}^{2}=0$. \emph{}The work of
Ref. shows that for the state $|\psi_{N}\rangle$, $\langle\hat{C}\rangle$
is nonzero. The experimental observation of the violation of (\ref{eq:sigcatone-2-2})
would signify failure of \emph{all} relevant classical mixture models,
and for a given $N$ is a Type I signature of the $N$-scopic cat-state.
\begin{figure}

\includegraphics[width=0.8\columnwidth]{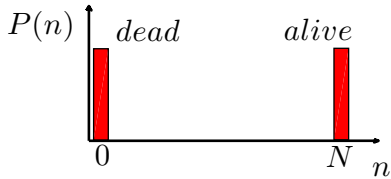}

\vspace{-1cm}

\caption{\emph{Signifying a cat-state by falsifying the quantum realism model
(\ref{eq:mixtureI}).} Measurements $\hat{M}$ on a system yield binary
outputs.  The Type I signature negates the model (\ref{eq:mixtureI})
for this system, where the quantum density operators $\rho_{1}$ and
$\rho_{2}$ give the ``dead'' and ``alive'' results respectively.}
\end{figure}

Similar considerations give Type I signatures for the two-mode NOON
superposition state 
\begin{equation}
|\psi_{NOON}\rangle=\frac{1}{\sqrt{2}}\Bigl(|N\rangle_{A}|0\rangle_{B}+|0\rangle_{A}|N\rangle_{B}\Bigr)\label{eq:supnoon}
\end{equation}
that has been prepared in the laboratory \cite{opticalNOON-1}. The
$\hat{a},$ $\hat{b}$ are boson destruction operators for two modes
denoted $A$ and $B$, respectively. The $|n\rangle_{A}$ is the eigenstate
of mode number $\hat{n}_{a}=\hat{a}^{\dagger}\hat{a}$ and similarly
$|n\rangle_{B}$ is the eigenstate of $\hat{n}_{b}=\hat{b}^{\dagger}\hat{b}$.
One can define $\hat{M}$ as the mode number difference $J_{z}=(\hat{n}_{a}-\hat{n}_{b})/2$.
The binned regions $1$ and $2$ are those that give outcomes for
$J_{z}$ as either negative or positive, respectively. Similar to
the above case (\ref{eq:sup1}), the NOON state $|\psi_{NOON}\rangle$
predicts a binary distribution as in Figure 1. As one example of a
Type I signature, the system prepared in a NOON state can be rigorously
distinguished from all classical mixtures (\ref{eq:mixtureI}) by
the observation of $\langle\hat{a}^{\dagger N}\hat{b}^{N}\rangle\neq0$
\cite{bognoon,murray}. This moment has been measured by higher order
interference fringe patterns as explained in Refs. \cite{opticalNOON-1,bognoon}.

We conclude this section by noting that most previous approaches for
signifying a Schrodinger cat-state use a Type I signature of some
sort (though sometimes with additional assumptions) e.g. see Refs.
\cite{ystbeamsplit-1-1,cats,cats2,catsphil}.

\section{Type II cat-signatures: Negating localised macroscopic hidden variable
state realism}

The next question is how to negate probabilistic classical mixtures
where the cat-system can be ``dead'' \emph{or} ``alive'', \emph{without}
the assumption that the component states of the mixtures are necessarily
quantum states. This question has been analysed in the literature,
but different analyses have introduced different extra assumptions
(e.g. we will compare Refs. \cite{LG,svet,MLR,mdrmlr2,eprcat,noonbellalex}).
In this Section, we examine signatures for the cat-state based on
the additional assumption of \emph{locality} between the cat-system
$C$ and a second remote system $S$.

\subsection{Localised macroscopic hidden variable states}

We consider Schrodinger's original formulation of the cat-paradox,
where the cat-system is entangled with a second system: A common example
is \cite{cats}
\begin{equation}
|\psi\rangle=\frac{1}{\sqrt{2}}\Bigl(|-\alpha\rangle_{C}|\downarrow\rangle_{S}+|\alpha\rangle_{C}|\uparrow\rangle_{S}\Bigr)\label{eq:catent}
\end{equation}
Here $|\uparrow\rangle$, $|\downarrow\rangle$ are the spin-$1/2$
eigenstates for $\hat{J}_{z}$ and the cat-system is modelled as the
single bosonic mode in a coherent state $|\alpha\rangle$. 

A second example is the Greenberger-Horne-Zeilinger (GHZ) state comprising
$N$ spin-$1/2$ particles \cite{svetcats,svet}: 
\begin{equation}
|\psi_{GHZ}\rangle=\frac{1}{\sqrt{2}}\Bigl(|\uparrow\rangle^{\otimes N}-|\downarrow\rangle^{\otimes N}\Bigr)\label{eq:ghz-1-1-3}
\end{equation}
This system can be divided into two subsystems and written as
\begin{equation}
|\psi_{GHZ}\rangle=\frac{1}{\sqrt{2}}\Bigl(|\uparrow\rangle_{C}^{\otimes N-k}|\uparrow\rangle_{S}^{\otimes k}-|\downarrow\rangle_{C}^{\otimes N-k}|\downarrow\rangle_{S}^{\otimes(k)}\Bigr)\label{eq:ghz-1-1}
\end{equation}
Here $|\uparrow\rangle^{\oplus N-k}=\prod_{m=1}^{N-k}|\uparrow\rangle^{(m)}$
and $|\uparrow\rangle^{\oplus k}=\prod_{m=N-k+1}^{N}|\uparrow\rangle^{(m)}$
where $|\uparrow\rangle^{(m)}$ is the spin eigenstate for $\hat{\sigma}_{Z}^{(m)}$,
the $\hat{\sigma}_{Z}$ observable for the $m$-th particle. The $|\downarrow\rangle^{\oplus N-k}$
and $|\downarrow\rangle^{\oplus k}$ are defined similarly in terms
of the eigenstates $|\downarrow\rangle$. The $\hat{\sigma}_{Z}$,
$\hat{\sigma}_{X}$ and $\hat{\sigma}_{Y}$ are the Pauli spin observables.
We classify the first $N-k$ particles as being part of the cat-system
$C$ and the remaining $k$ particles as forming the second system
denoted $S$. In this case, the measurement $\hat{M}$ is the collective
spin $\sum_{m=1}^{N-k}\hat{\sigma}_{Z}^{(m)}$ of the cat-system $C$
and the ``dead'' and ``alive'' outcomes symbolised in Figure 1
correspond to the results $N-k$ and $-(N-k)$.

Another example of an entangled cat-system is the NOON state (\ref{eq:supnoon})
where the mode $A$ is the cat-system $C$ and the mode $B$ is the
system $S$. Here, $\hat{M}=\hat{n}_{a}$ and the dead and alive
outcomes are numbers $0$ and $N$ as in Figure 1.

To describe a Schrodinger cat state \emph{without} the assumption
that the dead and alive states are quantum states, we assume a hidden
variable model in which the cat-system $C$ is always either in a
hidden variable state for which the cat is ``dead'',\emph{ or }in
a hidden variable state for which the cat is ``alive''. These two
hidden variable states need not be quantum states, which limits the
criteria that can be applied to negate such a model. For example the
Type 1 signature (\ref{eq:sigcatone-2-2}) that assumes the uncertainty
relation for each dead and alive state is no longer useful. In order
to derive suitable criteria, we introduce the further assumption of
\emph{locality} between the two systems $C$ and $S$ of the cat-states
(\ref{eq:catent})-(\ref{eq:ghz-1-1}) and (\ref{eq:supnoon}). We
call such dead and alive hidden variable states (subject to the assumption
of locality) \emph{localised macroscopic hidden variable states}. 

The locality assumption is based on the principle that the two subsystems,
the ``cat'' $C$ and the spin $S$, can become spatially separated,
so that measurements made on them can be space-like separated. The
assumption of local hidden variables states implies that the joint
probability for a result $x_{c}$ and $x_{s}$ upon measurements $X_{C}(\theta)$
and $X_{S}(\phi)$ on the cat and spin systems respectively can be
written in the form of Bell's local hidden variable model (LHV):
\begin{equation}
P(x_{c},x_{s})=\int\rho(\lambda)P_{C}(x_{c}|\theta,\lambda)P_{S}(x_{s}|\phi,\lambda)d\lambda\label{eq:lhv}
\end{equation}
Here the hidden variable state is given by a \emph{set of variables}
denoted $\lambda$ and $\rho(\lambda)$ is the associated probability
density. The $\theta$ and $\phi$ represent the choice of measurement
made at $C\equiv A$ and $S\equiv B$. The $P_{C}(x_{c}|\theta,\lambda)$
and $P_{S}(x_{s}|\phi,\lambda)$ are probabilities for the outcome
$x_{c}$ (or $x_{s}$) given the hidden state $\lambda$. The factorisation
in the integrand reflects the locality assumption that the probability
of the outcome at one site does not depend on the choice of measurement
made at the other site. 
\begin{figure}

\includegraphics[width=0.5\columnwidth]{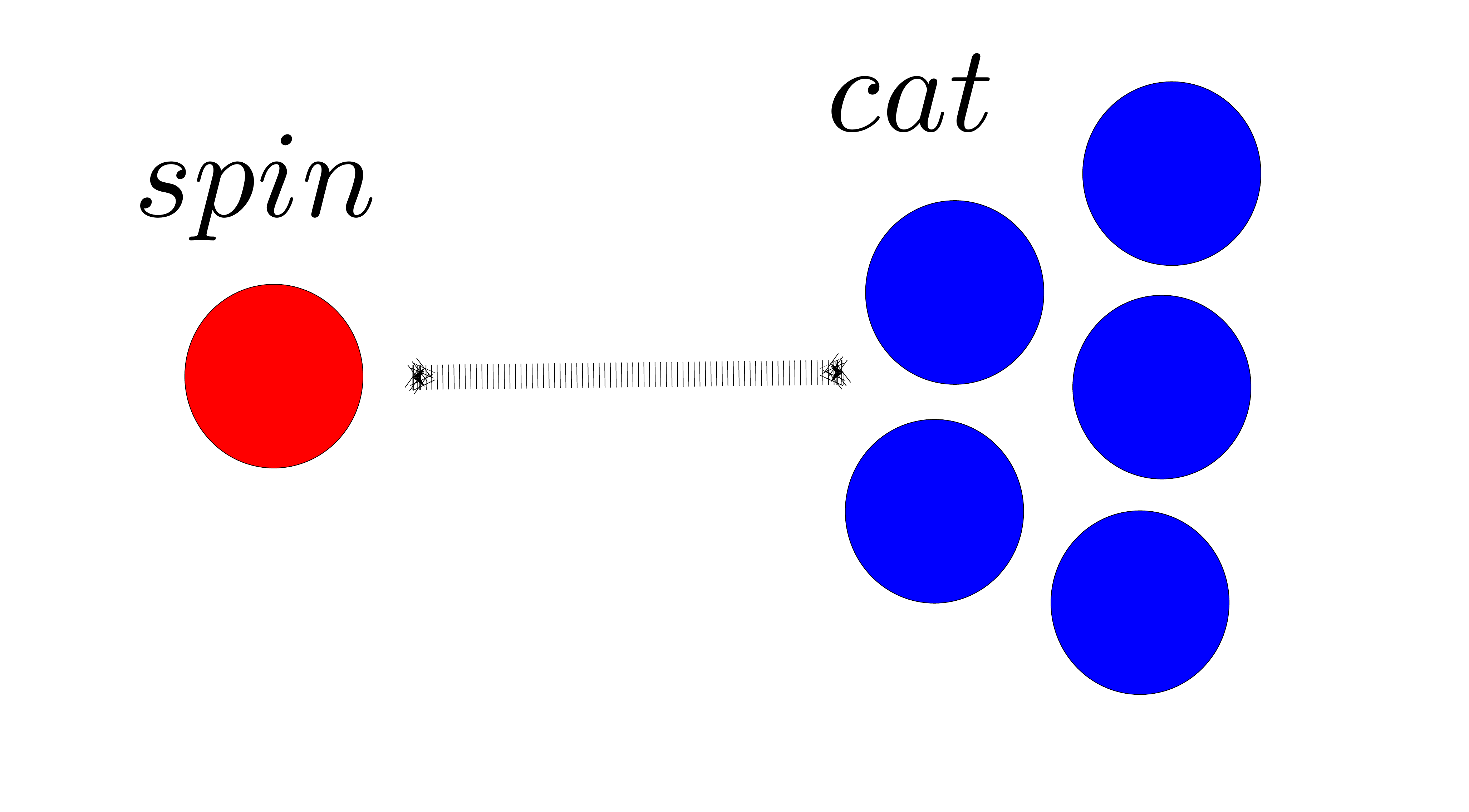}\includegraphics[width=0.5\columnwidth]{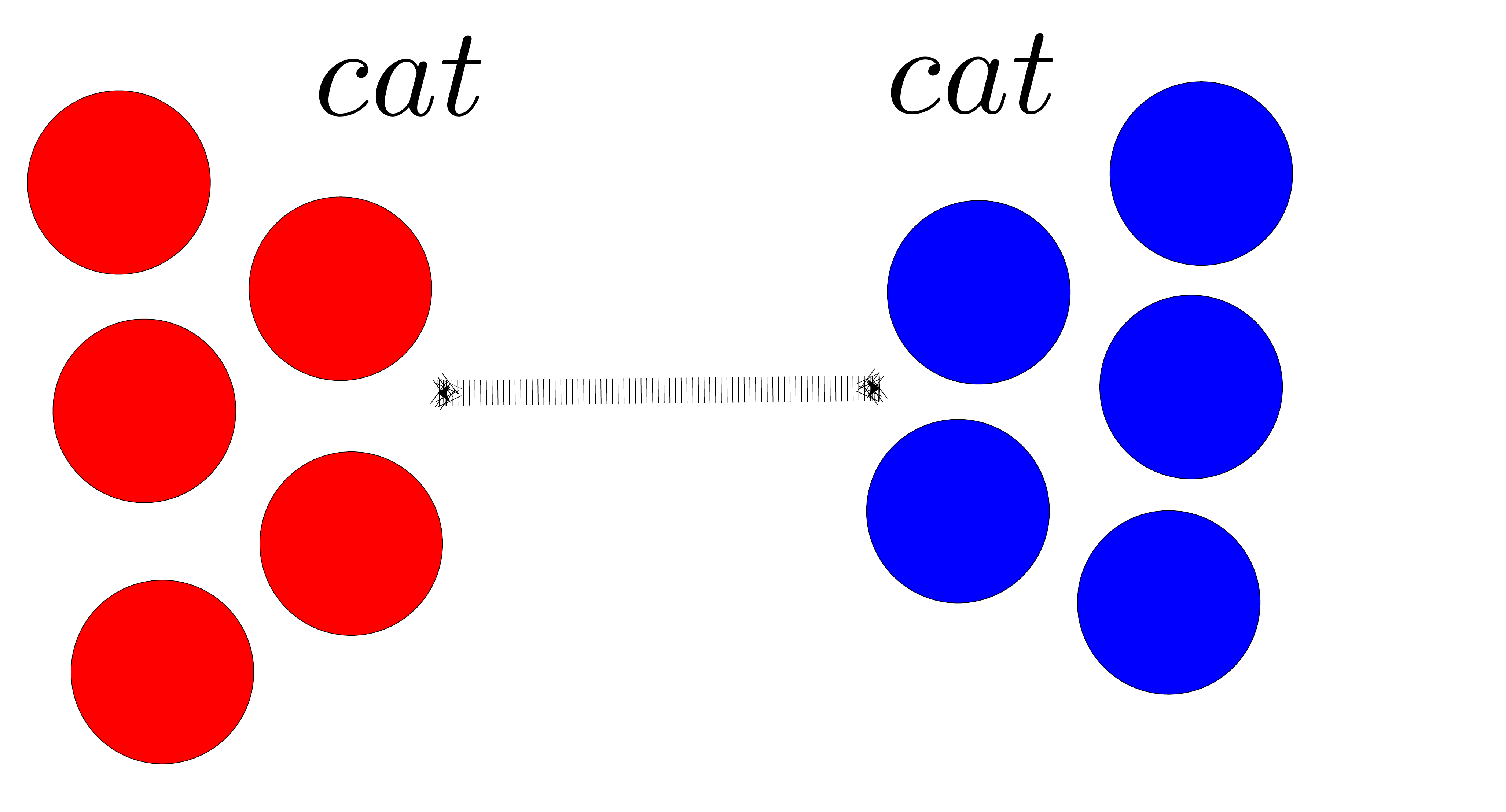}

\caption{\emph{Signifying a Schrodinger cat-system (\ref{eq:ghz-1-1}) by falsifying
all localised macroscopic hidden variable state (LMHVS) models (\ref{eq:lhvm}).}
Locality is assumed between the two subsystems $S$ (red) and $C$
(blue), but not within each subsystem.  The violation of the Svetlichny-type
multipartite Bell inequality falsifies all such models for the GHZ
state, including for the system of two cats (top right). }
\end{figure}

\subsection{The macroscopic hidden variable $\lambda_{M}$: Correlation, LHV
models and the macroscopic pointer}

The premise of macroscopic realism (MR) for the cat-system $C$ places
an additional restriction on the LHV model (\ref{eq:lhv}). For consistency
with MR, each hidden variable state $\lambda_{M}$ comprises a \emph{macroscopic
hidden variable}, $\lambda_{M}$, which takes the value $+1$ if the
cat-system is ``alive'', and $-1$ if the cat-system is ``dead''. 

However, in the specific examples of the cat-states (\ref{eq:catent}),
(\ref{eq:ghz-1-1}) and (\ref{eq:supnoon}), this condition does not\emph{
}have to be imposed because for these particular correlated states,
it arises naturally as a \emph{consequence} of the LHV assumption
(\ref{eq:lhv}).  In each case, there is a correlation between the
systems $C$ and $S$, so that a measurement on the system $S$ will
imply the outcome (whether ``alive'' or ``dead'') for the cat-system
$C$. For example, for the GHZ state (\ref{eq:ghz-1-1}) the value
of $\hat{M}$ can be inferred from the collective spin measurement
$\hat{O_{S}}=\sum_{m=N-k+1}^{N}\hat{\sigma}_{Z}^{(m)}$ of the system
$S$. Similarly for the NOON state, the value of $\hat{M}=\hat{n}_{a}$
can be inferred from measurement $\hat{O}_{S}=\hat{n}_{b}$ on $S$.
Consistency with the LHV model then \emph{imposes} the condition that
there be a macroscopic hidden variable $\lambda_{M}$, to denote that
the for the cat-system $C$, the measurement $\hat{M}$ has a predetermined
outcome i.e. that the ``cat'' is predetermined ``dead'' or ``alive''.
This result is proved in Ref. \cite{mdrcat} but is part of the original
analysis of ``elements of reality'' given by Einstein-Podolsky-Rosen
\cite{epr}. 

To remind us of the need for consistency with MR, we rewrite the LHV
model (\ref{eq:lhv}) as 
\begin{eqnarray}
P(x_{C},x_{S}) & = & \int\rho(\lambda,\lambda_{M})P_{C}(x_{c}|\theta,\lambda,\lambda_{M})\nonumber \\
 &  & \ \ \ \ P_{S}(x_{s}|\phi,\lambda,\lambda_{M})d\lambda d\lambda_{M}\label{eq:lhvm}
\end{eqnarray}
where we make the macroscopic hidden variable $\lambda_{M}$ explicit
in the notation. We call this model a \emph{localised macroscopic
hidden variable state} model (LMHVS). We also note that this model
for the quantum states (\ref{eq:catent}), (\ref{eq:ghz-1-1}) and
(\ref{eq:supnoon}) is a model for a \emph{quantum measurement }of
the system $S$. The second system $C$ (the cat) acts as the \emph{measurement
pointer} of a measurement apparatus that measures an observable $\hat{O}_{S}$
of $S$. This is because the result for $\hat{M}$ (which gives the
measured state of the ``cat'', whether ``dead'' or ``alive'')
indicates the result of the measurement of the observable $\hat{O}_{S}$
of the first system $S$. The association in the model (\ref{eq:lhvm})
of a macroscopic hidden variable $\lambda_{M}$ gives a theory in
which the macroscopic pointer is pointing ``either dead or alive''
at all times.

\subsection{Negating localised macroscopic hidden variable state realism}

The negation of the LMHVS model (\ref{eq:lhvm}) is possible using
certain Bell inequalities. To avoid the issue about which hidden variable
states are falsified (those of the cat system $C$ or the system $S$),
we consider the entangled cat-states where both systems $A\equiv C$
and $B\equiv S$ are large. Specifically, we consider the GHZ state
comprising $N$ spin-$1/2$ particles as two separated spin-systems
(Figure 2b) where both $k$ and $N-k$ are large. The negation of
the LHV model (\ref{eq:lhv}) for this system would tell us that there
can be no hidden variable state for each subsystem that is consistent
with locality between the two systems $C$ and $S$. In particular,
this negates that there can be any mixture for (at least one of) the
cat systems which enable the cat to be in a ``dead'' or ``alive''
\emph{local} state. 

For a system prepared in the GHZ state (\ref{eq:ghz-1-1}), the negation
of the LMHVS (\ref{eq:lhvm}) can be proved using Svetlichny Bell
inequalities derived in Refs. \cite{svet}. To summarise, consider
the complex operator $\Pi_{M}=\prod_{j=1}^{M}F_{j}$, $M\leq N$ where
$ $$F_{j}=\sigma_{X}^{(j)}+i\sigma_{Y}^{(j)}$ ($j\neq N$) at each
of the sites and $F_{N}=\sigma_{\pi/4}^{(N)}+i\sigma_{3\pi/4}^{(N)}$
($\sigma_{\theta}^{(j)}=\sigma_{X}^{(j)}\cos\theta+\sigma_{Y}^{(j)}\sin\theta$).
Observables $\mathrm{Re}\Pi_{M}$ and $\mathrm{Im}\Pi_{M}$ are defined
according to $\Pi_{M}=\mathrm{Re}\Pi_{M}+i\mathrm{Im}\Pi_{M}$. That
there cannot be a hidden variable set consistent the LMHVS model (\ref{eq:lhvm})
is proved using the fact that there are algebraic bounds $\langle\mathrm{Re}\Pi_{M}\rangle,\langle\mathrm{Im}\Pi_{M}\rangle\leq2^{M-1}$
and $\langle\mathrm{Re}\Pi_{M}\rangle+\langle\mathrm{Im}\Pi_{M}\rangle\leq2^{M}$
for any such underlying hidden variable state. The LHV model leads
to the Svetlichny-Bell inequality \cite{svet}
\begin{equation}
\langle\mathrm{Re}\Pi_{N}\rangle+\langle\mathrm{Im}\Pi_{N}\rangle\leq2^{N-1}\label{eq:svet}
\end{equation}
 The inequalities are predicted to be violated by the GHZ state,
which gives the prediction 
\begin{equation}
\langle\mathrm{Re}\Pi_{N}\rangle+\langle\mathrm{Im}\Pi_{N}\rangle=2^{N-1/2}\label{eq:qmpred}
\end{equation}
In fact the violation holds for all bipartitions (\ref{eq:ghz-1-1})
of the $N$ spin systems i.e. for all values of $k$. In this way,
we see that we negate any hidden variable model for the ``cat''
system of any size, conditional that the state be consistent with
the locality assumption between the two (potentially macroscopic)
systems $C$ and $S$. 

Other Bell inequalities have been constructed that could be applied
to negate the LMHVS model for the NOON state (\ref{eq:supnoon}) and
the entangled cat-state \cite{noonbellalex}
\begin{equation}
|\psi\rangle=\frac{1}{\sqrt{2}}\{|-\alpha\rangle|-\alpha\rangle+|\alpha\rangle|\alpha\rangle\}\label{eq:catent-1}
\end{equation}
These violations have been tested in some experimental situations
\cite{bellcatexp}. Such violations demonstrate the failure of any
hidden variable state to describe a cat-system, given that this state
must be consistent with the assumption of locality between the cat-system
$C$ and a second system $S$.

\section{Interpreting the Type I and II cat-signatures}

\subsection{Microscopic effects}

If a Type I or II cat-signature is observed in an experiment, then
it cannot be ruled out that the signature is due to a\emph{ microscopic}
quantum effect. This is because the Type I and II cat-signatures involve
predictions for measurements other than\emph{ }the macroscopic measurement
$\hat{M}$. In order to signify the cat-state using these signatures,
it is necessary to make assumptions about the \emph{microscopic} predictions
for these measurements $-$ for example that they are consistent with
locality down to a single atom or photon level. The consequence is
that the Type I and II cat-signatures are not sufficient to negate
the validity of the \emph{macroscopic} hidden variable $\lambda_{M}$
\cite{mdrcat}.

To illustrate, the Type I signature given by $\langle\hat{a}^{\dagger N}\hat{b}^{N}\rangle\neq0$
for the NOON state is observable as an interference pattern with frequency
proportional to $N$ (see Refs. \cite{opticalNOON-1}). The pattern
is increasingly difficult to resolve as $N\rightarrow\infty$ e.g.
\emph{all} photons need to be detected at either one location or another.
(See Ref. \cite{gisinuncert} for more general results). 

Similarly, the Type II signature given by the violation of the Svetlichny
Bell inequality (\ref{eq:svet}) requires measurement of the spins
$\hat{\sigma}_{X}$, $\hat{\sigma}_{Y}$ of \emph{all} the $N$ particles.
To negate the hidden variable model (\ref{eq:lhvm}), it is therefore
necessary to assume that the hidden variable states $\lambda$ give
predictions for microscopic features of the cat-system. Bounds
on the detail required to signify certain cat-states by a Type II
signature have been given in Ref. \cite{MLR-uncert,mdrcat} where
it is shown that a measurement resolution at the quantum noise level
is necessary.

To summarise: The Type I and II signatures of the cat-state are \emph{a
negation that the cat is in an alive or dead state, where the meaning
of ``state'' is that the ``state'' gives microscopic details in
the predictions of measurements made on the cat-system}. e.g. the
Type II signatures are a negation of the hidden variable \emph{states}
that give microscopic detail in the predictions. Thus, if we signify
the cat-state, we can only say that the cat is \emph{neither in a
dead state, nor in an alive state,} where the ``state'' means a
description of the system that gives microscopic details of certain
measurements.

\subsection{Macroscopic pointer}

Bearing in mind that the macroscopic hidden variable \emph{predetermines}
the result for the macroscopic measurement, if we cannot negate the
macroscopic hidden variable $\lambda_{M}$, then the simplest interpretation
is that the cat was indeed ``dead'' or ``alive'' prior to the
measurement, and that the signature of the cat-state is evidence of
failure of the assumptions made about the microscopic predictions
for the system \cite{mdrcat}. 

For example, it cannot be excluded that the signatures are due to
a \emph{microscopic nonlocal} effect. The LHV model (\ref{eq:lhvm})
assumes \emph{full locality} between the two systems $C$ and $S$.
If this full locality is relaxed by a small amount (to allow small
changes of size $\delta$ in the cat-state due to measurements on
the spin), then the signature of the cat-state is nullified. 

The interpretation is depicted in Figure 3. Here, the macroscopic
pointer does indeed point to one of two macroscopically distinct locations
$x_{1}$ and $x_{2}$ on a measurement dial. In terms of the state
(\ref{eq:catent}), the positions represent the $\hat{x}$ outcomes
$x_{1}$ and $x_{2}$ corresponding to the coherent states $|\alpha\rangle$
or $|-\alpha\rangle$ respectively ($\alpha$ is real). The positions
are not defined with a microscopic precision, however, and the pointer
may have an indeterminacy $\delta$ in position/ momenta. This is
associated with a potential nonlocal effect of size $\delta$.

In the context of many cat-signatures (see Refs. \cite{mdrcat,mdrmlr2,MLR-uncert,ystbeamsplit-1-1}),
``microscopic'' implies a size $\delta$ of an order defined by
the Heisenberg uncertainty bound. In the above, the addition of noise
$\delta_{x}$ and $\delta_{p}$ to the measurements of $x$ and $p$
where $\delta_{x}\delta_{p}\sim1/2$ is known to destroy the quantum
effect $-$namely, the signature of the cat \cite{MLR-uncert}.
\begin{figure}

\ \ 

\includegraphics[width=0.5\columnwidth]{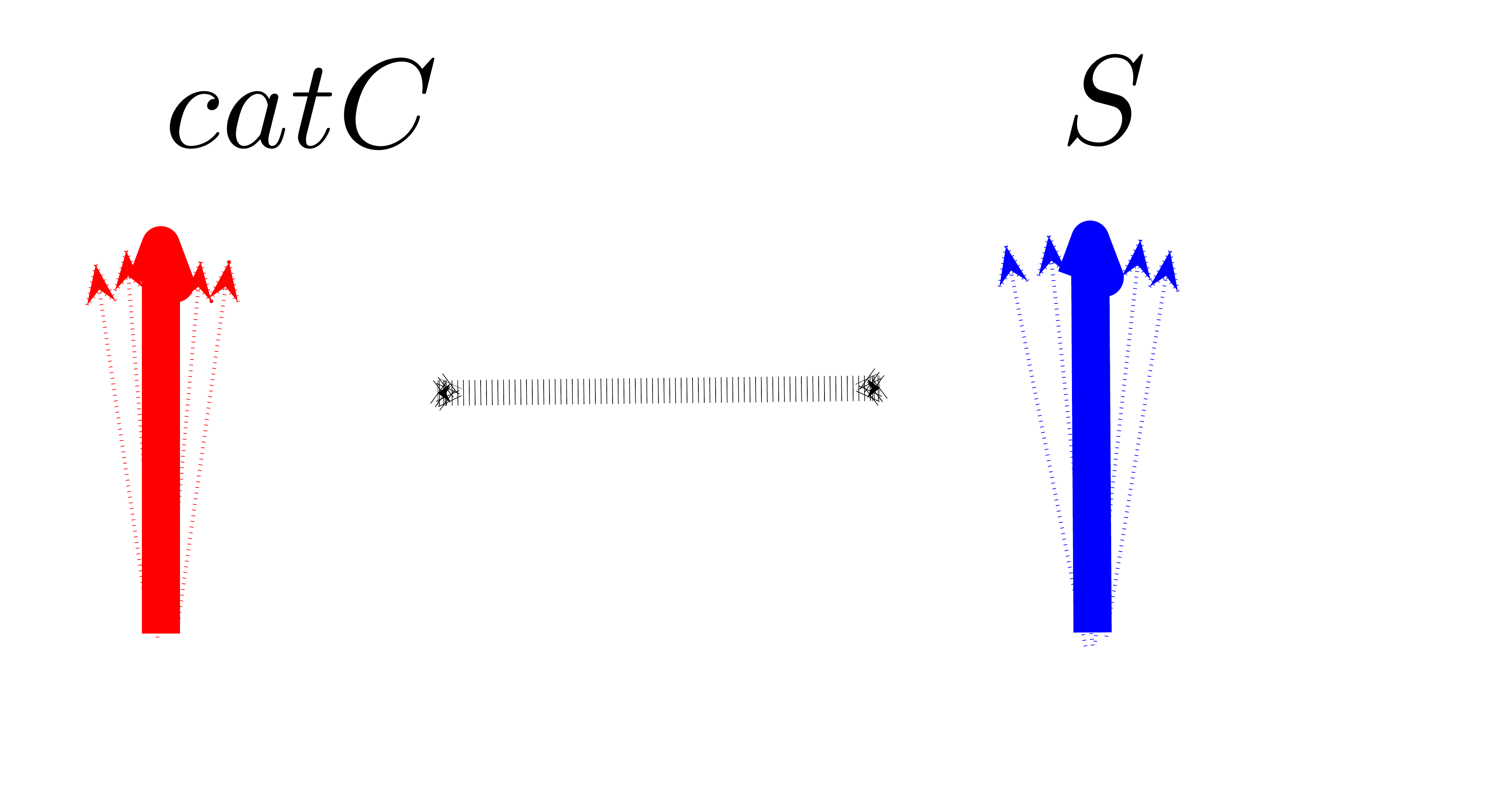}\includegraphics[width=0.5\columnwidth]{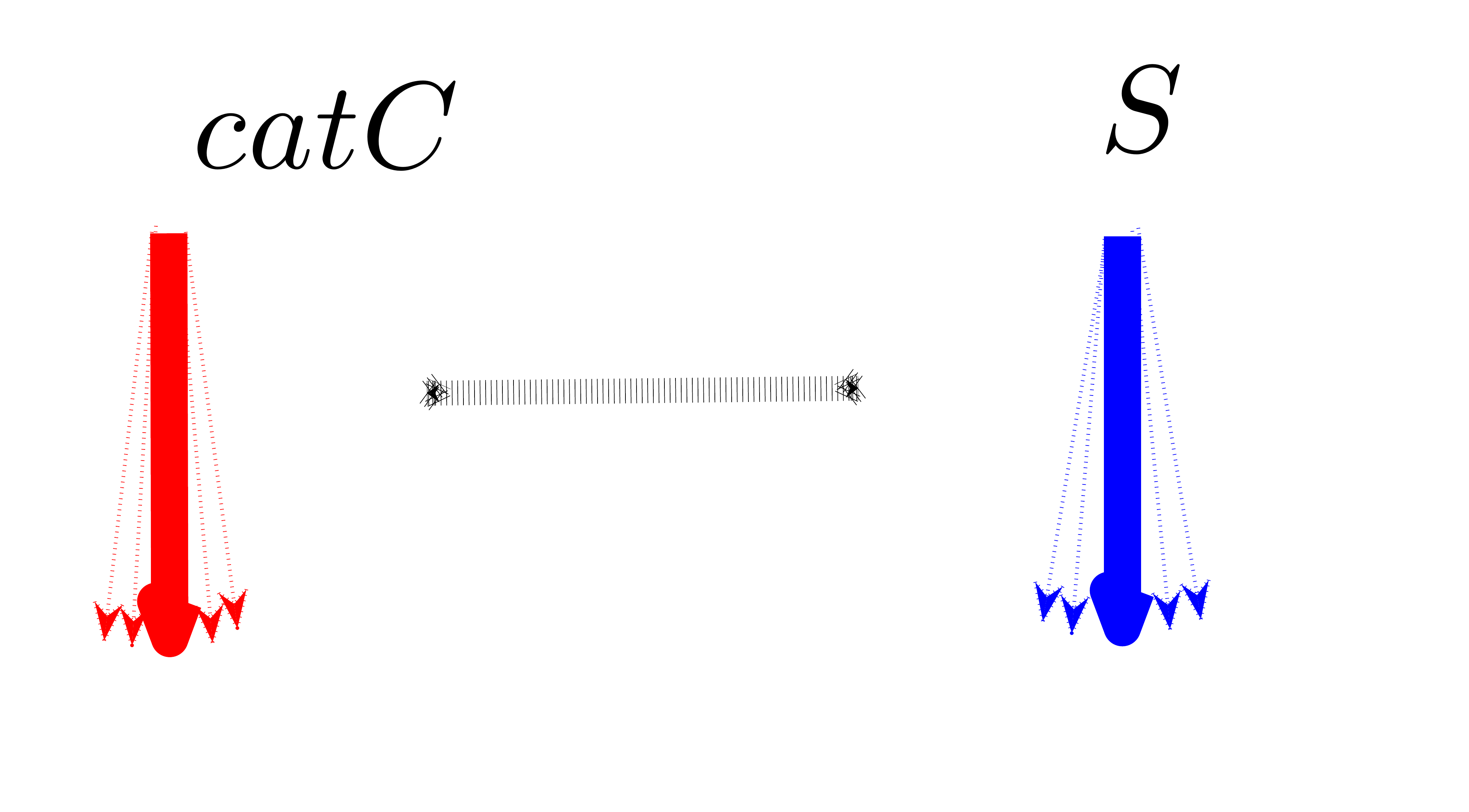}

\caption{\emph{Macroscopic pointer:} The pointer in a quantum measurement process
is modelled as the cat-system $C$ of the cat-state (\ref{eq:cat}).
The system being measured is modelled as the spin-system $S$. When
a measurement of the spin $\sigma_{Z}$ occurs, there are two final
positions for the pointer. While a possible quantum interpretation
of the cat-state is that the pointer is ``simultaneously in both
positions'', the Type I and II signatures of the cat-state cannot
negate the macroscopic hidden variable $\lambda_{M}$ that predetermines
the pointer to be at one location or the other. It cannot be excluded
that the Type II signature is due to microscopic nonlocal effects
of size $\sim\delta$ between the pointer $C$ and the spin $S$ (lower
diagram). The cat-state is consistent with the pointer being in one
position or the other (though with a microscopic indeterminacy associated
with the microscopic nonlocality).}
\end{figure}

\section{Type III cat-signatures: Negating macroscopic local realism}

A strong way to signify a Schrodinger cat-state is to falsify the
macroscopic hidden variable $\lambda_{M}$. Since this hidden variable
is a predetermination of the macroscopic measurement $M$ \emph{only}
(not other measurements), the falsification of this variable would
imply a genuine negation of ``\emph{macroscopic reality} (MR)''.
In that case, one can say the ``cat'' is neither dead nor alive,
where this means the measurement outcome for $\hat{M}$ is not predetermined,
in analogy with  interpretation discusssed in Schrodinger's essay.
There have been proposals to falsify MR by negating the macroscopic
hidden variable $\lambda_{M}$, a well-known example being the Leggett-Garg
proposal \cite{LG}. This proposal however involves a second premise.
Logically, wherever a second premise is introduced, it is necessary
to examine the second premise closely, since a signature can occur
\emph{if the second premise fails, with the first premise (macroscopic
realism) being upheld.} 

In this Section, we examine an alternative test of macroscopic realism,
one in which macroscopic realism is defined in conjunction with the
second premise of \emph{macroscopic locality}. This means that we
consider two spatially separated systems $A$ and $B$, and spacelike
separated measurements made on each one. The combined premise we refer
to as \emph{macroscopic local realism} (MLR). We argue that the premise
of macroscopic locality is a suitable co-premise of macroscopic realism,
in that the falsification of MLR is as significant as falsification
of MR.

\subsection{Bell Inequalities for MLR}

The premise of MLR combines the premises of \emph{macroscopic realism}
and \emph{macroscopic locality}. \emph{Macroscopic realism} is that
the system $A$ (or $B$) is in one of two macroscopically distinguishable
states at all time, in the sense of the macroscopic hidden variable
$\lambda_{M}^{A}$ (or $\lambda_{M}^{B}$) being predetermined. It
is thus assumed that a measurement $\hat{M}^{A}$ made on system $A$
reads out the value of the hidden variable $\lambda_{M}^{A}$, defined
with a macroscopic degree of fuzziness; and similarly for a measurement
$\hat{M}^{B}$ at $B$. \emph{Macroscopic locality} is that the measurement
$\hat{M}$ on one system cannot bring about an immediate macroscopic
change to the system at the other location. By a macroscopic change
in this context, we mean a transition of the macroscopic hidden variable
$\lambda_{M}$ being $+1$ to being $-1$ or vice versa i.e. a transition
between ``dead'' and ``alive''. The premise of macroscopic locality
asserts that a measurement cannot make a \emph{macroscopic} change
to another system, but we cannot exclude that it can make a \emph{microscopic}
one. The premise is therefore less strict than the premise of locality
(or local realism) which excludes all changes, microscopic and macroscopic,
and which has been negated. 

Let us consider two spatially separated systems $A$ and $B$ and
spacelike separated measurements $\hat{M}_{\theta}^{A}$ and $\hat{M}_{\phi}^{B}$
that can be made on each system. Here $\theta$ and $\phi$ are measurement
settings and we consider two measurement choices $\theta$, $\theta'$
and $\phi$, $\phi'$ for each system. We suppose that the measurements
$\hat{M}_{\theta}^{A}$, $\hat{M}_{\theta'}^{A}$ and $\hat{M}_{\phi'}^{B}$,
$\hat{M}_{\phi}^{B}$ each give macroscopically distinct binary outcomes
which are denoted $+1$ and $-1$ (corresponding to ``alive'' and
``dead'' regimes 2 and 1 shown in Figure 4). If we assume macroscopic
local realism, the following CHSH Bell inequality will hold \cite{MLR}
\begin{eqnarray}
\langle\hat{M}_{\theta}^{A}\hat{M}_{\phi}^{B}\rangle-\langle\hat{M}_{\theta}^{A}\hat{M}_{\phi'}^{B}\rangle+\langle\hat{M}_{\theta'}^{A}\hat{M}_{\phi}^{B}\rangle+\langle\hat{M}_{\theta'}^{A}\hat{M}_{\phi'}^{B}\rangle & \leq & 2\nonumber \\
\label{eq:bell-1}
\end{eqnarray}
The MLR model is an example of an LHV model and the derivation of
(\ref{eq:bell-1}) is therefore that of the standard CHSH Bell inequality
that applies to all LHV models where the measurements have binary
outcomes \cite{CHSHBell}. The violation of (\ref{eq:bell-1}) will
imply failure of MLR. Violations of Bell inequalities for cat-states
have been predicted and observed experimentally \cite{bellcatexp,noonbellalex,svetcats}.
However these do not involve macroscopic outcomes for all measurements
$\theta$, $\theta'$, $\phi$ and $\phi'$ and hence do not violate
(\ref{eq:bell-1}). That signatures of a cat-state require at least
one measurement to be finely resolved is a generic property discussed
in Refs \cite{gisinuncert,mdrcat,ystbeamsplit-1-1}. This would seem
to make the violation of (\ref{eq:bell-1}) impossible.
\begin{figure}

\includegraphics[width=0.8\columnwidth]{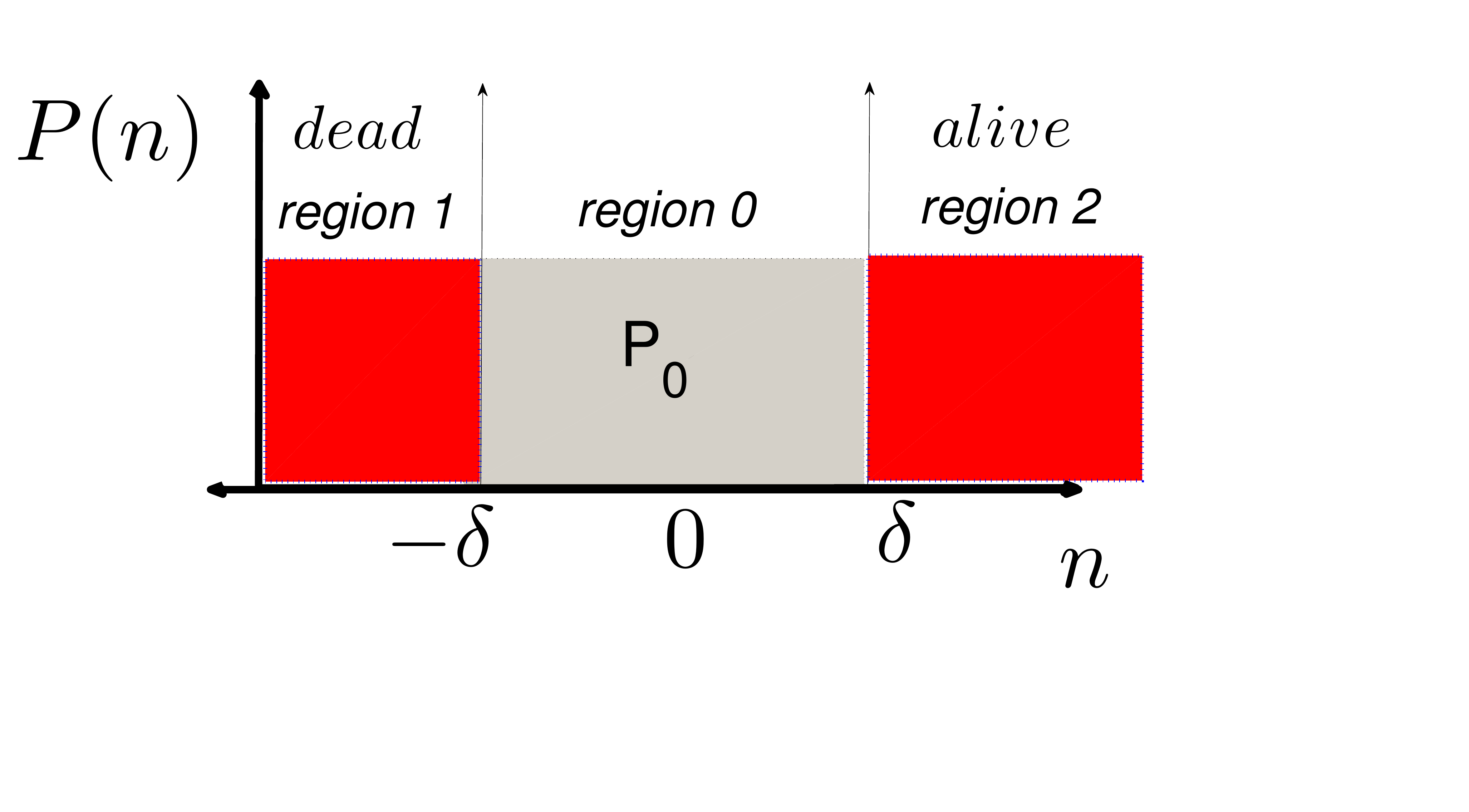}

\vspace{-1cm}

\caption{\emph{Practical method for testing $\delta$-scopic LR.} The outcomes
of each measurement indicated in Figure 5 is binned into one of the
regions $1,0,2$. As $\alpha\rightarrow\infty$, $P_{0}\rightarrow0$.}
\end{figure}

As might be expected, however, the possibility of violating the inequality
(\ref{eq:bell-1}) depends on how we interpret ``\emph{macroscopic}''.
First, we generalise the definition of MLR by defining $\delta$-scopic
local realism ($\delta$-LR). The $\delta$-scopic LR is falsified
where the separation between the outcomes for the measurements $\hat{M}_{\theta}^{A}$,
$\hat{M}_{\theta'}^{A}$ and $\hat{M}_{\phi'}$, $\hat{M}_{\phi}^{B}$
is greater than or equal to $2\delta$ (Figure 4). We next examine
scenarios where it may be possible to falsify $\delta$-scopic local
realism for some quantifiable $\delta$ that can be made large by
an amplification process that involves measurement of quantum noise.
In the scenarios that we consider, the amplifcation process occurs
as part of a measurement process, similar to the Schrodinger-cat gedanken
experiment.

\subsection{Amplification of the quantum noise level}

We now consider in detail proposals that have been put forward for
violating $\delta$-scopic local realism using field quadrature phase
amplitude observables. The crucial point is that measurement of the
field amplitudes takes place via an amplification process that involves
a second field, so that the final measurement is of a Schwinger spin
\cite{mdrmlr2,MLR}. The uncertainty principle for spin is
\begin{equation}
\Delta\hat{J}_{X}^{A}\Delta\hat{J}_{Y}^{A}\geq|\langle\hat{J}_{Z}^{A}\rangle|/2\label{eq:urs}
\end{equation}
One is able to create a situation where the quantum noise level given
by $|\langle\hat{J}_{Z}^{A}\rangle|/2$ is amplified to a very large
photon number difference (field intensity). This allows consideration
of changes of order $\delta$ where $\delta$ is large in the absolute
sense of particle number (intensity) but small compared to the quantum
noise level. The highly non-classical mesoscopic effects that are
predicted can then be understood as a property of amplified quantum
fluctuations.

The system we consider comprises two spatially separated modes at
$A$ and $B$ (Figure 5). We denote the modes at $A$ and $B$ by
the boson operators, $\hat{a}_{1}$ and $\hat{b}_{1}$, respectively.
At each location, the mode $\hat{a}_{1}$ (or $\hat{b}_{1}$) is combined
with a second mode $\hat{a}_{2}$ (or $\hat{b}_{2}$) respectively.
This combination can occur through a $50/50$ beam splitter. The outputs
at each location are rotated modes with boson operators given as $\hat{c}_{+}=(\hat{a}_{1}+\hat{a}_{2})\sqrt{2}$
and $\hat{c}_{-}=(-\hat{a}_{1}+\hat{a}_{2})/\sqrt{2}$ for $A$, and
$\hat{d}_{+}=(\hat{b}_{1}+\hat{b}_{2})/\sqrt{2}$ and $\hat{d}_{-}=(-\hat{b}_{1}+\hat{b}_{2})/\sqrt{2}$
for $B$. At each location, an experimentalist makes a measurement
of a number difference $N_{+}-N_{-}$ defined 
\begin{equation}
\hat{J}_{\theta}^{A}(\varphi)=(N_{+}-N_{-})/2=(\hat{c}_{2}^{\dagger}\hat{c}_{2}-\hat{c}_{1}^{\dagger}\hat{c}_{1})/2\label{eq:spinmr}
\end{equation}
where $\hat{c}_{2}=\hat{c}{}_{+}\cos\theta+e^{i\varphi}\hat{c}_{1}\sin\theta$
and $\hat{c}_{1}=-\hat{c}_{+}\sin\theta+e^{i\varphi}\hat{c}_{-}\cos\theta$.
This could be carried out using a phase shift $\varphi$ and polarising
beam splitters rotated to $\theta$ with the modes $c_{+}$ and $c_{-}$
as inputs. The measurement (\ref{eq:spinmr}) corresponds to a measurement
of the Schwinger spin observables $\hat{J}_{X}^{A}$, $\hat{J}_{Y}^{A}$,
$\hat{J}_{Z}^{A}$ at $A$ for the operators $a_{1}$ and $a_{2}$.
\begin{eqnarray}
\hat{J}_{X}^{A}=\hat{J}_{0}^{A}(\varphi) & = & (\hat{a}_{2}^{\dagger}\hat{a}_{1}+\hat{a}_{1}^{\dagger}\hat{a}_{2})/2\nonumber \\
J_{Y}^{A}=\hat{J}_{\pi/4}(\pi/2) & = & (\hat{a}_{2}^{\dagger}\hat{a}_{1}-\hat{a}_{1}^{\dagger}\hat{a}_{2})/(2i)\nonumber \\
\hat{J}_{Z}=\hat{J}_{\pi/4}(0) & = & (\hat{a}_{2}^{\dagger}\hat{a}_{2}-\hat{a}_{1}^{\dagger}\hat{a}_{1})/2\label{eq:spinsch}
\end{eqnarray}
The spin observables at $B$ are defined similarly as
\begin{equation}
\hat{J}_{\phi}^{B}(\gamma)=(\hat{d}_{2}^{\dagger}\hat{d}_{2}-\hat{d}_{1}^{\dagger}\hat{d}_{1})/2\label{eq:spinB}
\end{equation}
where $\hat{d}_{2}=\hat{d}{}_{+}\cos\phi+e^{i\gamma}\hat{d}{}_{-}\sin\phi$
and $\hat{d}_{1}=-\hat{d}_{+}\sin\varphi+e^{i\gamma}\hat{d}_{-}\cos\varphi$.
We define the Schwinger observables at $B$ as $\hat{J}_{X}^{B}=(\hat{b}_{2}^{\dagger}\hat{b}_{1}+\hat{b}_{1}^{\dagger}\hat{b}_{2})/2$,
$\hat{J}_{Y}^{B}=(\hat{b}_{2}^{\dagger}\hat{b}_{1}-\hat{b}_{1}^{\dagger}\hat{b}_{2})/(2i)$
and $\hat{J}_{Z}^{B}=(\hat{b}_{2}^{\dagger}\hat{b}_{2}-\hat{b}_{1}^{\dagger}\hat{b}_{1})/2$.
\begin{figure}

\includegraphics[width=1\columnwidth]{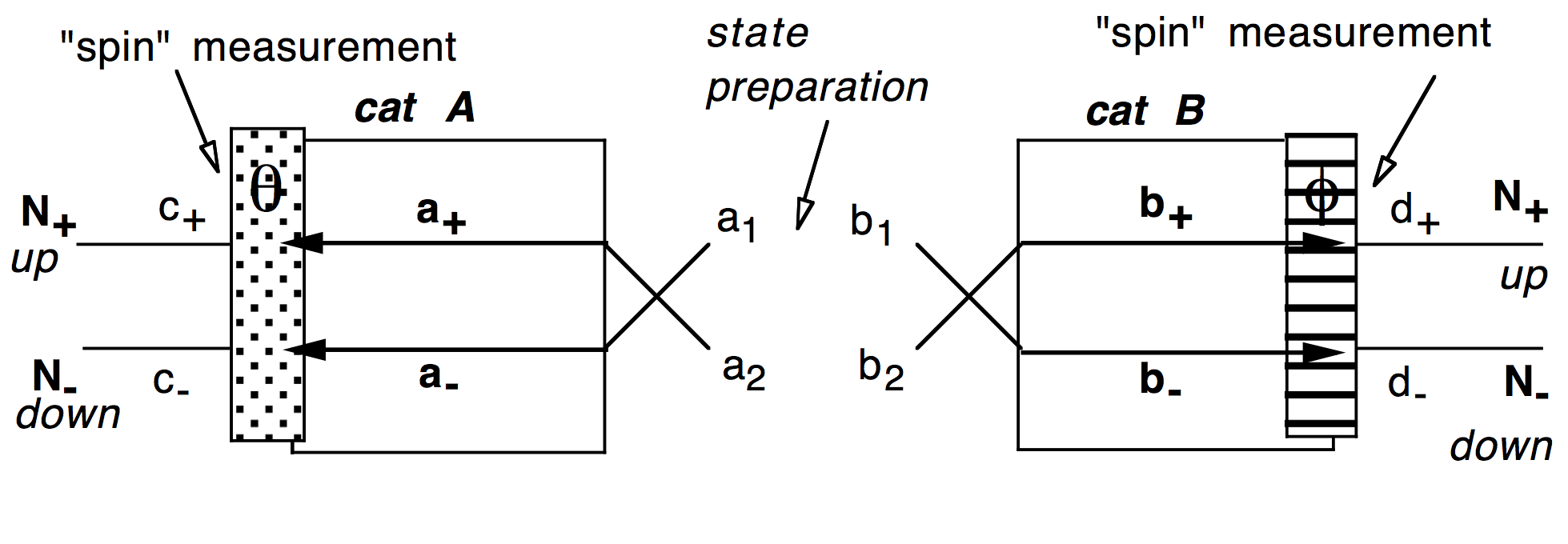}

\caption{\emph{Two Schrodinger cat-type states created as part of the measurement
process:} The modes $a_{1}$ and $b_{1}$ are created in an entangled
state $|\psi\rangle$. The final measurement of the number difference
$J=(N_{+}-N_{-})/2$ at each detector is an ``amplified'' value
of the quadrature phase amplitudes $x_{\theta}$ and $x_{\phi}$ of
the fields $a_{1}$ and $b_{1}$. The amplifcation is created at the
beam splitter where the fields $a_{2}$ and $b_{2}$ are independent
intense coherent states $|\alpha\rangle$. A Schrodinger cat-type
system is then created at each site $A$ and $B$.}
\end{figure}

Experiments have been performed where the modes $a_{1}$ and $b_{1}$
are created in an entangled state and the fields $a_{2}$ and $b_{2}$
are (to a good approximation) intense classical fields of amplitude
$\alpha$ (which we take to be real), similar to local oscillator
fields \cite{polsqueezing,polsqother}. Thus, each of the modes $c_{\pm}$
prior to the polarisation measurement $\hat{J}_{\theta}^{A}(\varphi)$
has (potentially) a \emph{macroscopic }photon number (and similarly
for the fields at $B$). In the experiments, a final polarisation
entanglement between the fields at $A$ and $B$ is signified via
measurements of $\hat{J}_{\theta}^{A}(\varphi)$ and $\hat{J}_{\phi}^{B}(\gamma)$.
The measurements $\hat{J}_{\theta}^{A}(\varphi)$ and $\hat{J}_{\phi}^{B}(\gamma)$
are also measurements of the\emph{ }quadrature phase amplitudes $\hat{x},\hat{p}$
of the original modes $a_{1}$ and $b_{1}$. This is because we can
simplify:
\begin{eqnarray}
\hat{J}_{X}^{A}=\hat{J}_{0}^{A}(\pi/2) & = & \alpha(a_{1}+a_{1}^{\dagger})/2=\alpha\sqrt{2}\hat{x}^{A}\nonumber \\
\hat{J}_{Y}^{A}=\hat{J}_{\pi/4}(\pi/2) & = & \alpha(a_{1}-a_{1}^{\dagger})/(2i)=\alpha\sqrt{2}\hat{p}^{A}\nonumber \\
\hat{J}_{Z}=\hat{J}_{\pi/4}(0) & = & \alpha^{2}/2\label{eq:spinhomo}
\end{eqnarray}
where $\hat{x}^{A}=(\hat{a}_{1}^{\dagger}+\hat{a}_{1})/\sqrt{2}$
and \textcolor{black}{$\hat{p}^{A}=i(a_{1}^{\dagger}-a_{1})/\sqrt{2}$}.
The Heisenberg uncertainty relation is $\Delta\hat{x}^{A}\Delta\hat{p}^{A}\geq1/2$.
A similar result holds for the quadrature phase amplitudes $\hat{x}^{B}=(\hat{b}_{1}^{\dagger}+\hat{b}_{1})/\sqrt{2}$
and \textcolor{black}{$\hat{p}^{B}=i(\hat{b}_{1}^{\dagger}-\hat{b}_{1})/\sqrt{2}$}
defined at $B$. In fact $\hat{J}_{\theta}(\pi/2)=\alpha\sqrt{2}\hat{x}_{2\theta}^{A}$
where $\hat{x}_{\theta}=\hat{x}\cos\theta+\hat{p}\sin\theta$.

We envisage an experiment where at site $A$, the experimentalist
can measure either $\hat{J}_{X}^{A}=\hat{J}_{0}^{A}(\pi/2)$ or $\hat{J}_{Y}^{A}=\hat{J}_{\pi/4}(\pi/2)$.
In terms of the original fields, using the result (\ref{eq:spinhomo}),
this corresponds to either $\alpha\sqrt{2}\hat{x}^{A}$ or $\alpha\sqrt{2}\hat{p}^{A}$.
Each of $\hat{J}_{X}^{A}$ and $\hat{J}_{Y}^{A}$ is a measurement
of a particle number difference according to the expression (\ref{eq:spinB}).
The choice of whether to measure $\hat{J}_{X}$ or $\hat{J}_{Y}$
is made \emph{after} the combination of the mode $a_{1}$ with the
strong field $a_{2}$. The $\hat{J}_{X}^{A}$ and $\hat{J}_{Y}^{A}$
are thus measurements of the amplified quadrature phase amplitudes
$\alpha\sqrt{2}\hat{x}^{A}$ and $\alpha\sqrt{2}\hat{p}^{A}$. Similar
measurements are made at $B$, where one would measure either $\alpha\sqrt{2}\hat{x}^{A}$
or $\alpha\sqrt{2}\hat{p}^{A}$. Hence if one considers a change $\delta_{X}$
(or $\delta_{P}$) in the quadrature phase amplitude $\hat{X}$ (or
$\hat{P}$), one can define in this context an\emph{ }amplified change
$\delta=\alpha\sqrt{2}\delta_{X}$ (or $\alpha\sqrt{2}\delta_{P}$)
for the particle number difference measured by $\hat{J}_{X}$ (or
$\hat{J}_{Y}$). The change can be made \emph{arbitrarily large},
in an absolute sense, by increasing $\alpha$. 

We note the increase in $\alpha$ also amplifies the total number
of particles at each site (this being determined by $|\alpha|^{2}$).
The nature of the amplification is evident by the uncertainty relation
(\ref{eq:urs}) for the actual spin measurements which reduces in
this case to 
\begin{equation}
\Delta\hat{J}_{X}^{A}\Delta\hat{J}_{Y}^{A}\geq|\alpha|^{2}/4\label{eq:uncert}
\end{equation}
since $\alpha$ is taken to be very large. The amplification that
is crucial to creating the macroscopic states at the locations $A$
and $B$ is also an amplification of the quantum noise level, and
there is no amplification relative to this level. 

\subsection{Using states that violate Continuous Variable Bell inequalities}

One can now design experiments that are predicted to falsify a $\delta$-scopic
local realism. For some states, the correlations obtained for the
quadrature phase amplitude measurements $\hat{x}_{\theta}^{A}$ and
$\hat{x}_{\phi}^{B}$ at each site are predicted to violate a Bell
inequality. The outcome $x$ for the measurement $\hat{x}$ at each
site can be binned into regions of positive and negative values. We
define an observable $\hat{S}_{\theta}^{A}$ whose value is $+1$
if $x_{\theta}^{A}\geq0$ and $-1$ otherwise. A similar observable
$\hat{S}_{\phi}^{B}$ is defined at $B$, based on the quadrature
phase amplitude $\hat{x}_{\phi}^{B}$. It has been shown that for
certain states $|\psi\rangle$ and for certain angles $\phi$, $\phi'$,
$\theta'$ and $\theta$, the following Bell inequality is violated
\begin{equation}
E=\langle S_{\theta}^{A}S_{\phi}^{B}\rangle-\langle S_{\theta}^{A}S_{\phi'}^{B}\rangle+\langle S_{\theta'}^{A}S_{\phi}^{B}\rangle+\langle S_{\theta'}^{A}S_{\phi'}^{B}\rangle\leq2\label{eq:bell}
\end{equation}
thus negating the possibility of an LHV model describing the results
of those measurements. Since we can also write $\hat{J}_{\theta}^{A}=\alpha\sqrt{2}\hat{x}_{2\theta}^{A}$
and $\hat{J}_{\phi}^{B}=\alpha\sqrt{2}\hat{x}_{2\phi}^{B}$, this
inequality is also violated if we define $\hat{S}_{\theta}^{A}$ as
the observable with value $+1$ if $J_{\theta}^{A}\geq0$ and $-1$
otherwise and $\hat{S}_{\phi}^{B}$ as the observable with value $+1$
if $J_{\phi}^{B}\geq0$ and $-1$ otherwise. The violation implies
that there is no predetermined (local) hidden variable description
for the sign of the number differences $J_{\theta}^{A}$, $J_{\phi}^{B}$.
This has been pointed out in the Ref. \cite{mdrmlr2}. Because we
can amplify $\alpha$, this gives a situation whereby one can falsify
local hidden variables for measurements of particle number difference
that can tolerate an uncertainty (or poor resolution) that increases
as $\alpha$ increases, the uncertainty becoming macroscopic as $\alpha\rightarrow\infty$.
An example of the state $|\psi\rangle$ is the pair coherent state
\begin{equation}
|\psi\rangle=\frac{e^{r_{0}^{2}}}{2\pi\sqrt{I_{0}(2r_{0}^{2})}}\int_{0}^{2\pi}|r_{0}e^{i\zeta}\rangle|r_{0}e^{-i\zeta}\rangle d\zeta\label{eq:pair}
\end{equation}
($I_{0}$ is the modifed Bessel function, $r_{0}=1.1$) that is generated
near the threshold of nondegenerate parametric oscillation \cite{pair-coh}.
\begin{figure}

\hspace{-1cm}\includegraphics[width=0.6\columnwidth]{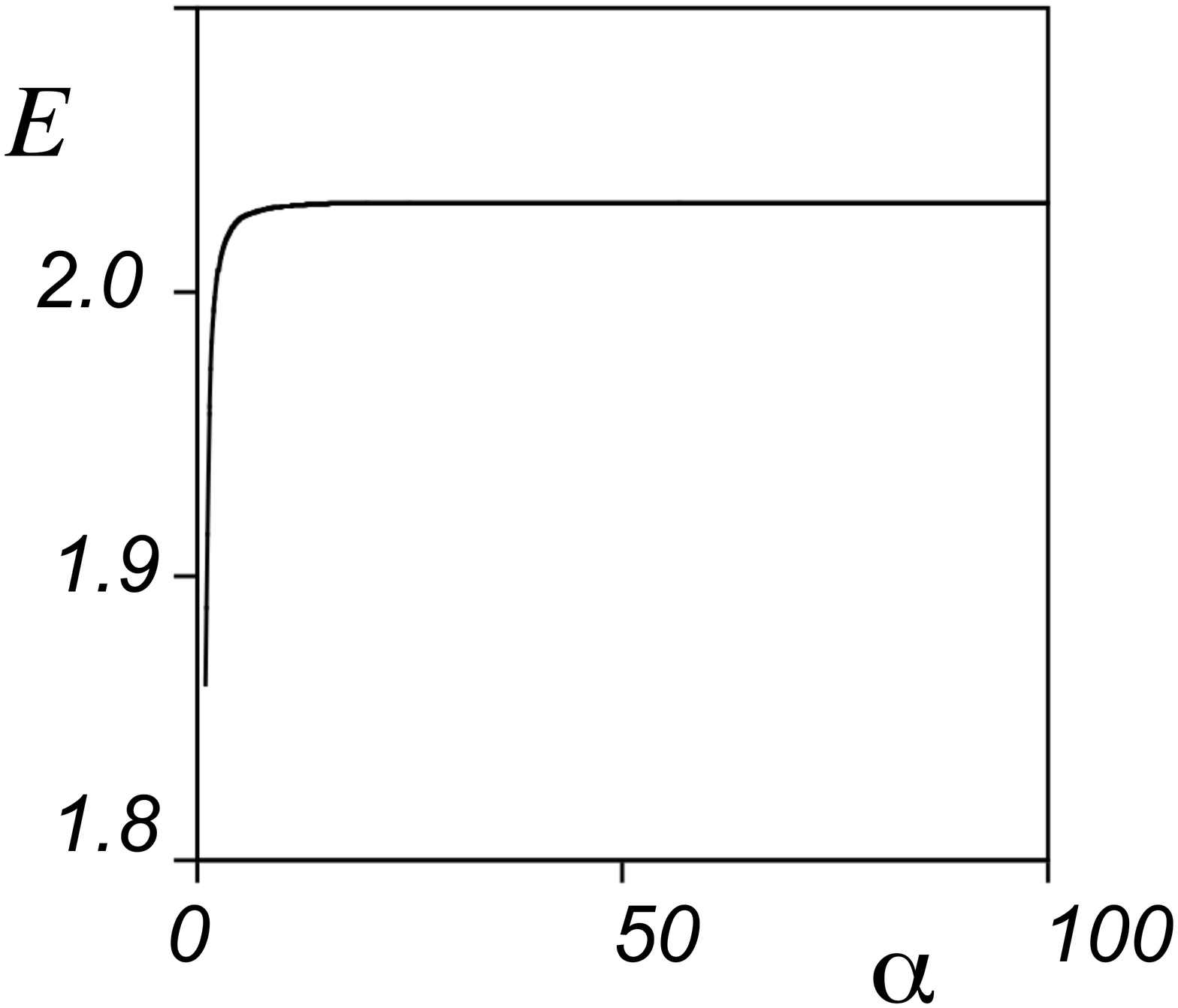}\negthickspace{}\negthickspace{}\negthickspace{}\negthickspace{}\negthickspace{}\negthickspace{}\negthickspace{}\negthickspace{}\negthickspace{}\negthickspace{}\includegraphics[width=0.6\columnwidth]{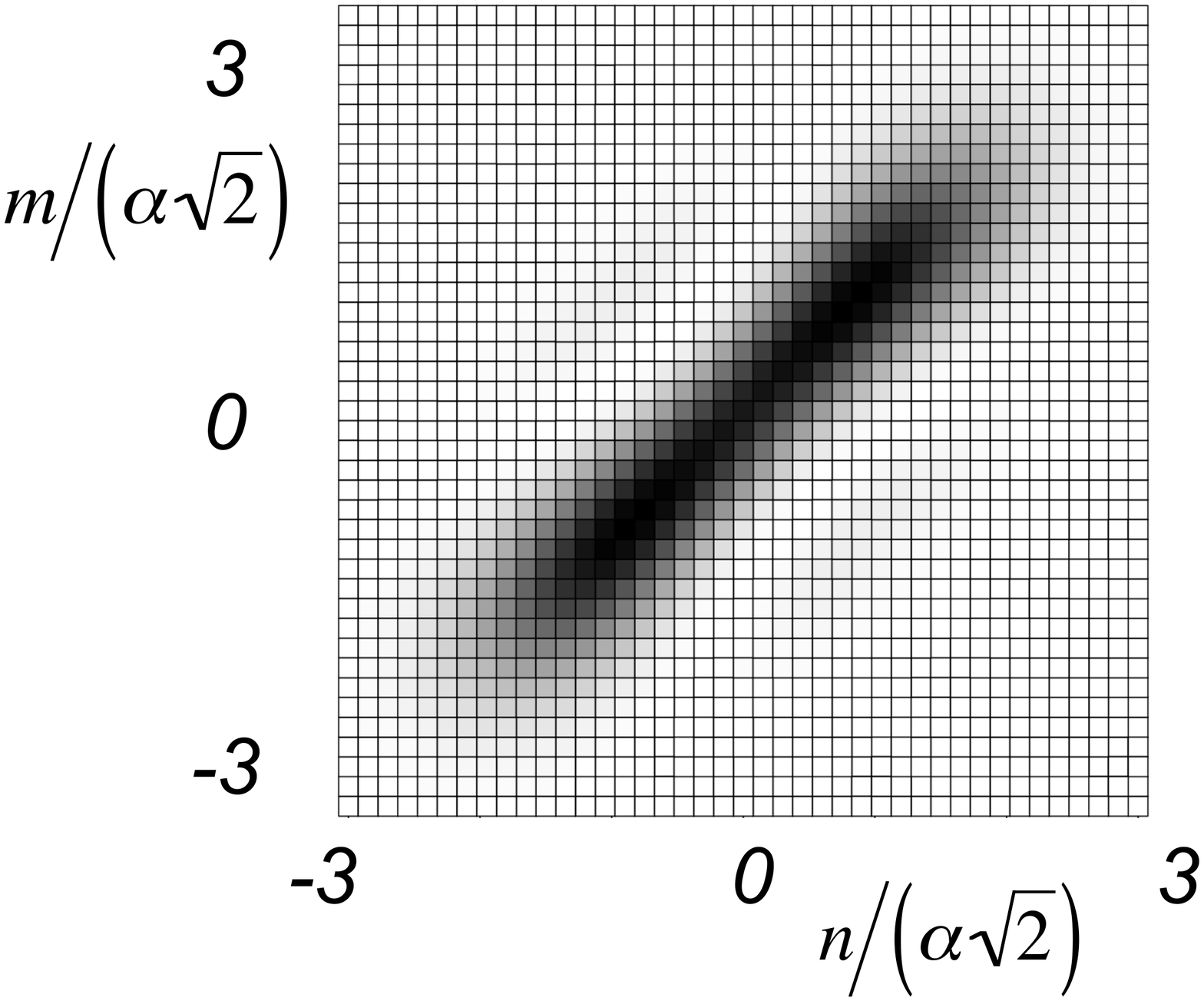}

\caption{\emph{Signature of the Schrodinger cat state created by apparatus
of Figure 4:} The number differences $J=(N_{+}-N_{-})/2$ at the site
$A$ and $B$ are denoted $n$ and $m$ respectively, and are binned
to the values $S=1$ or $-1$ according to sign as described in text.
Left: The expectation values $E$ violate the Bell inequality (\ref{eq:bell})
for all $\alpha\rightarrow\infty$. Right: A contour graph of the
probability for joint outputs $n$ and $m$. The absolute values of
the number difference outputs $n$, $m$ increase with $\alpha$.
}
\end{figure}

As $\alpha$ increases, we argue that the $+1$ and $-1$ outcomes
for $\hat{S}_{\theta}^{A}$ ultimately become macroscopically distinct
(and similarly the $+1$, $-1$ outcomes for $\hat{S}_{\phi}^{B}$
become macroscopically distinct). The measurements $\hat{S}_{\theta}^{A}$
and $\hat{S}_{\phi}^{B}$ are then examples of macroscopic measurements
$\hat{M}_{\theta}^{A}$ and $\hat{M}_{\phi}^{B}$ and the violation
of (\ref{eq:bell}) is a violation of (\ref{eq:bell-1}). In this
limit we would violate ``macroscopic local realism''. 

To understand the argument, we define a region of measurement outcome
$x$ for $\hat{J}_{\theta}^{A}$ where the result falls between $-\delta$
and $+\delta$ for some $\delta\neq0$ (see Figure 4). We call this
region $0$, and also define the region of outcome $x\geq\delta$
as region $2$, and the region of outcome $x\leq-\delta$ as region
$1$. Then for fixed $\delta$, the probability $P_{0}$ of a result
in the region $0$ becomes zero as $\alpha\rightarrow\infty$. Yet
the violation of the Bell inequality is unchanged with $\alpha$ (Figure
6). Hence, violation of the inequality (\ref{eq:bell-1}) is possible
for the two outcomes $+1$ and $-1$ that for large enough $\alpha$
can be justified as separated by a region of width $2\delta$. This
is true for any arbitrarily large fixed $\delta$, because $\alpha$
can be made larger without altering the Bell violation. Hence, there
is a prediction for a violation of mesoscopic/ macroscopic local realism. 

The violation of the inequality (\ref{eq:bell}) would imply a violation
of $\delta-$scopic local realism where $2\delta$ is the separation
between the outcomes $+$ and $-1$. For a realisation of the experiment,
however, there will be a small nonzero probability for a result in
the region $0$ and this must be taken into account. A method for
doing this is explained in the next section.

\subsection{Practical quantifiable $\delta$-scopic local realism tests }

The macroscopic realism premise (MR) would apply if $\delta$ is
macroscopic and $P_{0}=0$. Then MR asserts that if we consider two
states with outcomes confined to regions $1$ and $2$ respectively,
the system must be in a probabilistic mixture of those two states.
The meaning of MR for the more general case where $P_{0}\neq1$ is
discussed in the paper of Leggett and Garg \cite{LG} and further
in Refs. \cite{eric_marg,lauralg,bognoon}. 

The MR premise for this generalised case is that the system be described
as a probabilistic mixture of\emph{ }two\emph{ overlapping states}:
the first gives outcomes in regions ``$1$'' or ``$0$''; the
second gives outcomes in regions ``$0$'' or ``$2$''. The MR
assumption \emph{excludes} the possibility that the system can be
in a superposition of two states, one that gives outcomes in region
$1$ and the second that gives outcomes in $2$. It does not however
exclude superpositions of states with outcomes in region 1 and 0,
or superpositions of states with outcomes in regions 0 and 2. Where
$\delta$ is finite and not necessarily macroscopic, we use the term
$\delta$LR to describe the premise that is used. 

We follow the approach of Ref. \cite{lauralg}, and denote the hidden
variable state associated with the outcomes in regions ``$1$''
or ``$0$'' for the system at $A$ by the variable $\tilde{S}^{A}=-1$
and the hidden variable state that generates outcomes in regions ``0''
and ``2'' by $\tilde{S}^{A}=1$. We define the variable $\tilde{S}^{B}$
similarly. The macroscopic locality assumption applies to assert that
the measurement at one location cannot change the result at the other
in such a way that the system changes value of $\tilde{S}$ from $+1$
to $-1$, vice versa. We can define $P_{+}$ and $P_{-}$ as the probability
that the system is in the state with $\tilde{S}=+1$ or the other
state with $\tilde{S}=-1$. Then we note that the $\delta$-LR assumptions
would predict the Bell inequality
\begin{eqnarray}
E & = & \langle\tilde{S}_{\theta}^{A}\tilde{S}_{\phi}^{B}\rangle-\langle\tilde{S}_{\theta}^{A}\tilde{S}_{\phi'}^{B}\rangle+\langle\tilde{S}_{\theta'}^{A}\tilde{S}_{\phi}^{B}\rangle+\langle\tilde{S}_{\theta'}^{A}\tilde{S}_{\phi'}^{B}\rangle\leq2\nonumber \\
\label{eq:bell5}
\end{eqnarray}
\textcolor{black}{However, the }moments $K_{\theta\phi}=\langle\tilde{S}_{\theta}^{A}\tilde{S}_{\phi}^{B}\rangle$
are no longer directly measurable, because an outcome between $-\delta$
and $+\delta$ could arise from either state, $\tilde{S}=-1$ or $+1$.
However, we can always conclude that $P_{1}\leq P_{-}\leq P_{1}+P_{0}$
and $P_{2}\leq P_{+}\leq P_{2}+P_{0}$, where $P_{1},$ $P_{2}$ and
$P_{0}$ are the measurable probabilities of obtaining a result in
regions $1$, $2$ and $0$ respectively (Figure 4). Hence, we establish
bounds on the correlations assuming $\delta$LR, even if the $P_{0}$
are measured to have a nonzero probability.  The modified inequality
is 
\begin{equation}
E_{\delta}=K_{\theta,\phi}^{lower}-K_{\theta\phi'}^{upper}+K_{\theta'\phi}^{lower}+K_{\theta'\phi'}^{lower}\leq2\label{eq:ineq-1}
\end{equation}
where $K_{\theta\phi}^{lower}$ and $K_{\theta\phi}^{upper}$ are
lower and upper bounds to $K_{\theta\phi}$ i.e. $K_{\theta\phi}^{lower}\leq K_{\theta\phi}\leq K_{\theta\phi}^{upper}$.
We see that $K_{\theta\phi}^{lower}=P_{2,2}(\theta,\phi)+P_{1,1}(\theta,\phi)-P_{10,20}(\theta,\phi)-P_{20,10}(\theta,\phi)$
and $K_{\theta\phi}^{upper}=P_{20,20}(\theta,\phi)+P_{10,10}(\theta,\phi)-P_{1,2}(\theta,\phi)-P_{2,1}(\theta,\phi)$.
We introduce the notation that $P_{20,10}$, for example, is the joint
probability for an outcome $J_{z}$ in regions $2$ or $0$ at $A$
with the measurement angle set at $\theta$ and an outcome $x$ in
$1$ or $0$ at $B$ with the measurement angle set at $\phi$. 

The modified inequality (\ref{eq:ineq-1}) gives a practical means
to demonstrate a violation of an $\delta$-scopic local realism for
a finite $\delta$ where there is a small probability $P_{0}$ of
an outcome in the region defined by $-\delta<x<\delta$. A similar
inequality has been derived for Leggett-Garg experiments \cite{lauralg}.
For realistic tests based on current experiments, the shifts $\delta$
may not be macroscopic, but nonetheless offer a route to test local
realism beyond the single particle level considered in experiments
so far.
\begin{figure}

\includegraphics[width=0.8\columnwidth]{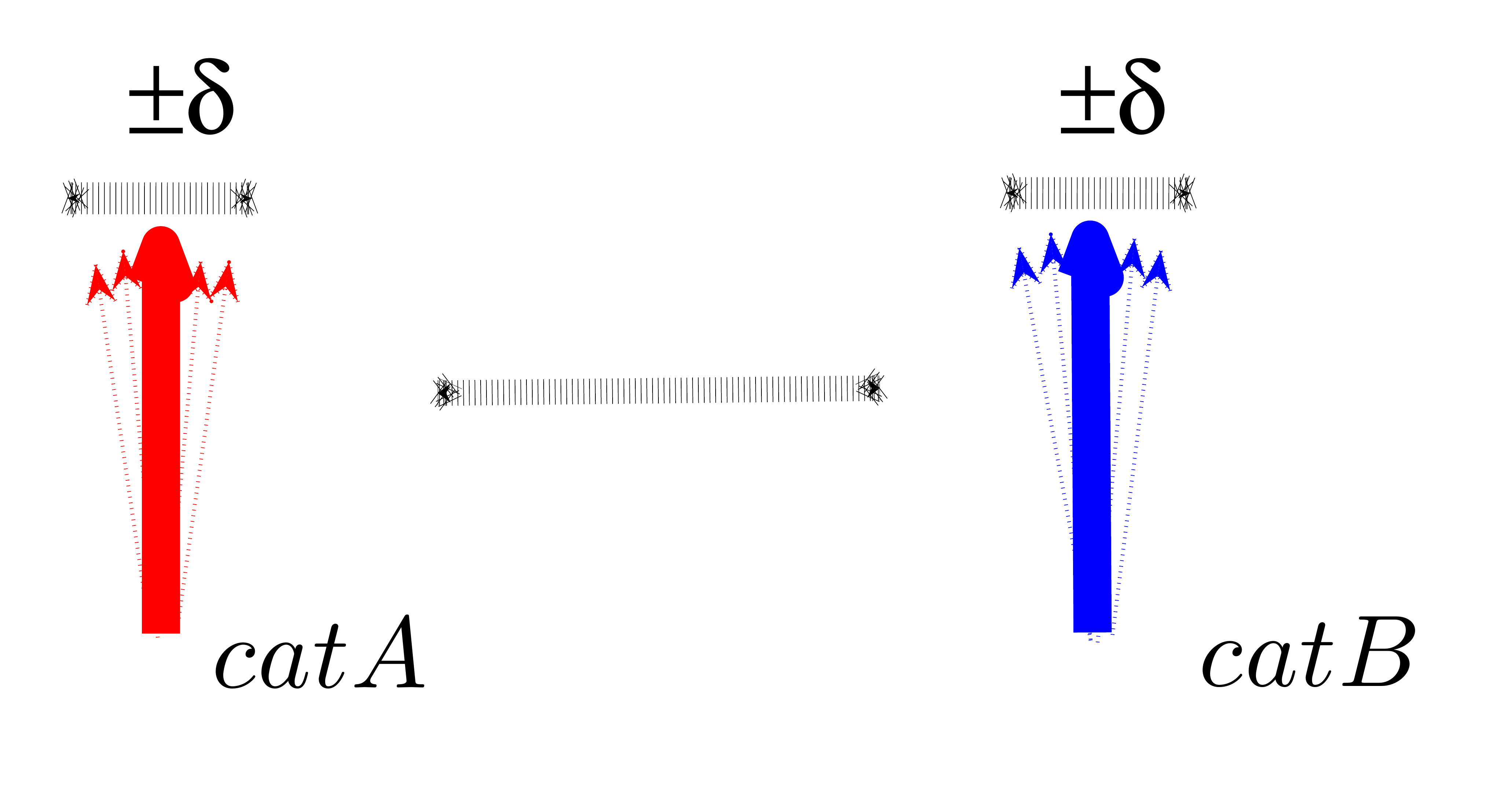}

\caption{\emph{Macroscopic pointers: }The macroscopic observables $J_{\theta}$,
$J_{\phi}$ for two cat-systems $A$ and $B$ depicted in Figure 5
are the macroscopic pointers for the quadrature phase amplitude measurements
of the microscopic systems $a_{1}$ and $b_{1}$. The violation of
the inequality (\ref{eq:ineq-1}) reveals a nonlocal effect of size
$\delta$ between the two pointers. This provides a test in which
the signature negates a genuine indeterminacy of the position of the
pointer, to within the $\pm\delta$. The size of $\delta$ can be
amplified by increasing the value of the coherent amplitude $\alpha$.
However, the indeterminacy is at the quantum noise level, and the
two pointers are not correlated at that level. }
\end{figure}

\subsection{The macroscopic pointers}

In the experiment of Figure 5, the two cat-states at $A$ and $B$
act as two pointers for the microscopic quadrature phase amplitudes
of the original entangled field modes denoted $a_{1}$ and $a_{2}$.
There is a correlation between the ``position'' $J$ of each pointer
as indicated by a particle number difference $N_{+}-N_{-}$ and the
original amplitude of the mode. However, the ``positions'' of the
two pointers are not well-correlated i.e. one pointer does not accurately
measure the position of the other, at least not to a precision given
by the quantum noise level of the uncertainty relation (\ref{eq:uncert}).
This is evident by the plot of Figure 6b which shows a weak correlation
between the quadrature phase amplitudes at each location. While the
pointers are entangled, they are not well-correlated: The range of
positions over which a pointer become interpretable as ``being in
simultaneously in both places'' (or else shifted between those two
places by measurements on a second pointer) is at this quantum noise
level.

\section{Discussion and Conclusion}

In summary we have examined different approaches to signifying a Schrodinger
cat-state, and contrasted with testing macroscopic realism. In Section
II we considered a model of a cat-system in which the cat is described
as a probabilistic mixture of two distinguishable \emph{quantum} states,
one describing the ``cat'' being ``dead'' and the other the ``cat''
being ``alive''. Criteria to negate this model (which we call \emph{macroscopic
quantum realism} MQR) were derived in the form of inequalities based
on the assumption that uncertainty relations hold for all quantum
states. We called this negation a Type I signature of a cat-state.

In Section III we examined models for the cat-system that do not require
the dead and alive states of the cat to be quantum states, but rather
allow them to be hidden variable states subject to the condition of
locality between the cat-system and a second remote system $S$. We
called this model a\emph{ localised macroscopic hidden variable state}
model (LMHVS). Criteria to negate the LMVS model were called Type
II signatures, and included the violation of multipartite Bell inequalities.

It was explained in Section IV that both the MQR and LMHVS models
make assumptions about \emph{microscopic} predictions for measurements.
Hence the Type I and Type II signatures do not directly falsify macroscopic
realism. Macroscopic realism (MR) asserts that the cat is \emph{predetermined}
dead or alive, prior to a coarse-grained measurement $\hat{M}$ that
distinguishes whether the cat is dead or alive (without measurement
of the other details of the system). Macroscopic realism therefore
asserts the validity of a macroscopic hidden variable $\lambda_{M}$
to describe the system: the $\lambda_{M}$ predetermines whether the
cat will be measured dead or alive according to a measurement $\hat{M}$.
Both the MQR and LMHVS models incorporate the macroscopic hidden variable
$\lambda_{M}$, but also assume other hidden variables that give a
predetermination for other measurements that are finely resolved.
We cannot therefore exclude that the results of an experiment signifying
the cat-state are caused by a microscopic nonlocal effect (such as
a change of spin of one of the particles in a GHZ state) rather than
a failure of MR.

In Section IV, we considered the classic example where the cat-system
$C$ models the macroscopic pointer of a measurement apparatus that
measures the spin of system $S$. After a measurement interaction,
quantum theory predicts the pointer $C$ to be entangled with the
system $S$. The entangled states are of the form of the cat-states
that we considered in Sections II and III. We argue that without the
negation of the macroscopic hidden variable of the pointer system,
the \emph{simplest }interpretation of the pointer is\emph{ not} that
it negates macroscopic realism (where the needle is pointing ``in
two places at once''). Rather, it can be interpreted that the pointer
is (approximately) at one place or the other but with small nonlocal
effects between the pointer $C$ and the measured system $S$.

The key question then becomes to find a scenario for testing macroscopic
realism where the observed effect cannot be explained by microscopic
nonlocality. We show in Section V how this might be possible provided
``macroscopically distinguishable outcomes'' refers to outcomes
with a large shift $\delta$ in particle number relative to two spatial
locations. For the examples that we consider however, the shift although
large in absolute terms is small relative to the total number of particles
of the system. Using this meaning of ``macroscopic'', we outline
a proposal to test \emph{macroscopic local realism} where two cat-systems
are generated using two entangled field modes prepared in a state
predicted to violate a continuous variable Bell inequality. A practical
method for testing mesoscopic local realism is outlined. The cat-systems
and the two macroscopically distinguishable outcomes for each cat-system
are created using an amplification brought about by local oscillator
fields. This amplification can be interpreted as part of the measurement
process, similar to Schrodinger's original example. In the proposed
experiments, the measurement process amplifies the microscopic quantum
noise levels into the more macroscopic fluctuations of a macroscopic
particle number difference observable. The highly non-classical mesoscopic
effects that are predicted can then be understood as a property of
amplified quantum fluctuations.
\begin{acknowledgments}
This work has been supported by the Australian Research Council under
Grant DP140104584.
\end{acknowledgments}

\end{document}